\documentclass[useAMS,usenatbib]{mn2e}

\usepackage{graphicx,amssymb,amsmath}

\title[The fraction of dark mass in two GCs]{Searching in the dark: the dark 
mass content of the Milky Way globular clusters NGC288 and NGC6218\thanks{Based on Fibre Large
Array Multi-Element Spectrograph (FLAMES) observations collected with the Very
Large Telescope of the European Southern Observatory, Cerro Paranal, Chile,
within the observing programmes 68.D-0212, 69.D-0582, 071.D-0131, 073.D-0211, 074.A-0508, 075.D-0043,
087.D-0276, 088.B-0403, 193.B-0936 and 193.D-0232 and on observations made 
with the NASA/ESA Hubble Space Telescope, which is operated by the association 
of Universities for Research in Astronomy, Inc., under the NASA contract 
NAS 5-26555, under programs GO-10775 (PI: Sarajedini).}}
\author[Sollima et al.]{A. Sollima$^{1,2}$\thanks{E-mail:
antonio.sollima@oabo.inaf.it}, F. R. Ferraro$^{2}$, L. Lovisi$^{2}$,
F. Contenta$^{3}$, E. Vesperini$^{4}$, L. Origlia$^{1}$,
\newauthor
E. Lapenna$^{2}$, B. Lanzoni$^{2}$, A. Mucciarelli$^{2}$, 
E. Dalessandro$^{2}$, C. Pallanca$^{2}$\\
$^{1}$ INAF Osservatorio Astronomico di Bologna, via Ranzani 1, Bologna, 40127,
Italy\\
$^{2}$ Dipartimento di Astronomia, Universit{\'a} di Bologna, via Ranzani 1,
40127, Bologna\\
$^{3}$ Department of Physics, University of Surrey, Guildford GU2 7XH, UK\\
$^{4}$ Department of Astronomy, Indiana University, Bloomington, IN 47405, USA
}
\begin{document}


\pagerange{\pageref{firstpage}--\pageref{lastpage}} \pubyear{2015}

\maketitle

\label{firstpage}

\begin{abstract}
We present an observational estimate of the fraction and distribution of dark mass in 
the innermost region of the two Galactic globular clusters NGC 6218 (M12) and
NGC 288. 
Such an assessment has been made by comparing the dynamical and luminous mass 
profiles derived from an accurate analysis of the most extensive spectroscopic 
and photometric surveys performed on these stellar systems.
We find that non-luminous matter constitutes more than 60\% of the total mass in
the region probed by our data ($R<1.\arcmin6\sim r_{h}$) in both clusters. We have 
carefully analyzed the effects of binaries and tidal heating on our estimate
and ruled out the possibility that our result is a spurious consequence of these
effects.
The dark component appears to be more concentrated than the most massive stars
suggesting that it is likely composed of dark remnants segregated in the
cluster core.
\end{abstract}


\section{Introduction}
\label{intro_sec}

The relative contribution of luminous and dark matter to the overall mass
budget of stellar systems contains crucial information on their
nature, origin and evolution. According to the $\Lambda CDM$ cosmological 
paradigm, the structures in the
Universe formed at high redshift through a
hierarchical assembly of small fragments of non-baryonic matter (White \&
Rees 1978). Galaxies of all morphological types are expected to form
within these fragments being nowdays embedded in dark matter
(DM) halos.
This evidence 
comes from the discrepancy between the mass estimated for these stellar systems 
using the kinematics of their stars and their luminosities as tracers. In
particular, the mass-to-light (M/L) ratios measured for these stellar systems range
from 5 (for dwarf elliptical galaxies) to $>$1000 (for ultra faint dwarf
spheroidals; Tollerud et al. 2011) i.e. several times larger
than those predicted by population synthesis models ($1.5<M/L<2.5$; Bruzual \&
Charlot 2003).

Globular clusters (GCs) appears to stand-out from this scenario. Indeed, at odds with
other DM dominated stellar systems populating contiguous regions of the 
luminosity-effective radius plane (Tolstoy, Hill \& Tosi 2009), they have low M/L ratios consistent with the
hypothesis they are
deprived of DM (McLaughlin \& van der Marel 2005; Strader, Caldwell \& Seth 2011).
This difference might suggest a different formation scenario for these stellar
systems which could form in collapsing DM-free gas clouds (see e.g. Kruijssen 2015). For this reason
the M/L ratios of GCs are often used as a reference for stellar population
studies and to validate the prediction of population synthesis models. 
On the other hand,
low-mass DM halos surrounding GCs progenitors are hypothesized by some model of GC
formation (Peebles 1984). These halos could be later stripped by the tidal interaction with the
host galaxy leaving only minimal imprints in the structural and kinematical 
properties of present day GCs (Mashchenko \& Sills 2005). Observational claims 
of the possible existence of DM dominated GCs in the giant elliptical NGC 5128 have been recently 
put forward by Taylor et al. (2015).

Non-baryonic DM is not the only invisible matter contained in stellar systems.
Indeed, the final outcome of the stellar evolution process of stars with 
different masses is represented by remnants (white dwarfs, neutron stars and
black holes) whose luminosities are comparable or even several orders
of magnitude smaller than those of the least 
luminous Main Sequence (MS) stars. The estimate of
the mass enclosed in dark remnants in a GC is complicated by the interplay
between stellar and dynamical evolution in these stellar systems and by the
uncertainties in their formation process (for a comprehensive discussion see
Heggie \& Hut 1996). 

Indeed, both neutron stars (NSs) and black
holes (BHs) are the compact remnants of massive ($M>8~M_{\odot}$) stars after
their explosion as SNe II. The off-center onset of the explosive mechanism can transmit 
to the remnant a velocity kick often exceeding the cluster escape speed thus
leading to its ejection outside the cluster (Drukier 1996; Moody \& Sigurdsson
2009).
Moreover, the large mass contrast of BHs with respect to the mean
cluster mass lead to a quick collapse of the BH population forming a 
dynamically decoupled sub-system in the central part of the cluster (Spitzer
1969). Scattering
between multiple BHs leads to a prompt ejection of these objects . 
For the above mentioned reasons 
it was initially suggested that GCs
should be deprived of BHs (Kulkarni, Hut \& McMillan 1993; Sigurdsson \&
Hernquist 1993). However recent observational studies (see e.g. Strader et al. 
2012; Chomiuk et al. 2013; Miller-Jones et al. 2015) have provided evidence of the
possible presence of stellar mass BHs in a number of GCs and several
theoretical studies (see e.g. Breen \& Heggie 2013; Heggie \& Giersz 2014; Morscher et al. 2013,
2015; Sippel \& Hurley 2013) have found that GCs might indeed still host a
non-negligible fraction of these compact objects. On the other hand, the large
fraction of X-ray 
binaries and pulsars per unit mass in GCs (up to a factor 100 larger than that
estimated in the Galactic field; Clark
1975; Grindlay \& Bailyn 1988) indicates that a
certain number of these objects must be present in these stellar systems.

White dwarfs (WDs) are the natural outcome of the evolution of low/intermediate
mass ($M<8~M_{\odot}$)
stars after the expulsion of their envelopes occurring at the end of their
asympthotic giant branch phase. As a consequence of the long lifetimes of WD
progenitors and the typical negative slope of 
the initial mass function (Kroupa 2001; Bastian, Covey \& Meyer 2010) the fraction of WDs 
steadly increases during the cluster lifetime making them a significant
contributor to the mass budget of a GC in the last stages of its evolution
(Vesperini \& Heggie 1997; Baumgardt \& Makino 2003).
Because of the particular form of the 
initial-final mass relation of low-mass stars (Kalirai et al. 2008) the mass 
spectrum of these objects is expected to be peaked at
$M\sim0.5~M_{\odot}$ i.e. only slightly larger than the present-day mean stellar
mass ($\sim0.4~M_{\odot}$)
and significantly smaller than the typical turn-off mass
in a GC ($\sim0.8~M_{\odot}$). So, at odds with NSs and BHs occupying always the high tail of the
mass distribution of stellar objects, stars evolving into WDs change their 
ranking in mass within the whole GC stellar population during their lifetimes. 
Consequently, two-body relaxation is 
expected to produce a progressive migration of WDs from the center, where their
massive progenitors sunk, towards the outer regions where stars with smaller
masses are preferentially located. Observational evidence of this phenomenon
seems to be provided by the analysis of the radial distribution of WDs in
different regions of the cooling sequence in 47 Tucanae
(Heyl et al. 2015). The interaction of a GC with the tidal field of its host
galaxy further complicate the prediction of the retention fraction of dark 
remnants. Indeed, the ever continuing injections of kinetic energy favors the
evaporation of the kinematically hottest (mainly low-mass) stars,
while massive objects (like NSs, BHs and massive WD progenitors) are
preferentially retained. N-body simulations indicate that the actual fraction 
of retained remnants has deep implications in the dynamical evolution of GCs
(Contenta, Varri \& Heggie 2015) and in their present-day M/L ratios
(L{\"u}tzgendorf, Baumgardt \& Kruijssen 2013).

In this context, many
observational analyses aimed at investigating the dark content in GCs
have been conducted in the past years by comparing the M/L ratios estimated from
stars kinematics and those predicted by stellar population
synthesis models (Richer \& Fahlman 1989; Meylan \& Mayor 1991; Leonard, Richer 
\& Fahlman 1992; Piatek et al. 1994; Dirsch \&
Richtler 1995; Ibata et al. 2013; L{\"u}tzgendorf et al. 2013; 
Kamann et al. 2014). One of the main drawbacks of this approach resides in the
choice of the uncertain parameters affecting the M/L like the present-day mass
function and the retention fraction of dark remnants (Shanahan \& Gieles 2015).

In Sollima et al. (2012) we determined the dynamical and
luminous masses of a sample of six Galactic GCs by fitting simultaneously their 
luminosity functions and their line-of-sight (LOS) velocity dispersion profiles
with multimass analytical models leaving the present-day mass function (MF) as a free
parameter and making an assumption on the dark remnants retention fraction. 
From this study we found that the derived 
stellar masses were systematically smaller than the dynamical ones by $\sim
40\%$. Although many hypotheses were put forward, the most favored interpretation
linked such a discrepancy to a fraction of retained dark
remnants larger than expected. Unfortunately, the robustness of the obtained result relies on the
ability of the adopted specific model in reproducing the actual
degree of mass segregation of the analysed clusters. In particular, both the
luminous and the dynamical masses of the best-fit model were constrained by
observables measured in the cluster core, a region where mass segregation
effects are maximized and where only a small fraction
of the cluster mass is contained, and then extrapolated to the whole cluster
(see Sollima et al. 2015).
Moreover, the approach adopted in that work did not allow to obtain information
on the radial distribution of the dark mass across the cluster, therefore
complicating any interpretation on its nature.

In this paper we present the result of a model-independent analysis of the most
extensive photometric and spectroscopic datasets available in the literature for 
two Galactic GCs, namely NGC 288 and NGC 6218 (M12) with the aim of deriving their dark mass content
and radial distribution. These objects are two well studied GCs located in the
southern emisphere which are particularly suited for this kind of studies. 
Both clusters are indeed relatively close ($d<9$ kpc) and characterized by a
small central projected density
($\Sigma_{V}>18~mag~arcsec^{-2}$; Harris 1996, 2010 edition) and for this reason 
it is possible to sample
their luminosity function and velocity dispersion profiles through photometric
and spectroscopic surveys even within their cores with no significant
crowding problems. In Sect. 2 we describe the observational dataset used in
the analysis and the data reduction techniques. In Sect. 3 the methods to derive
luminous and dynamical mass profiles for the two target clusters are described.
The fractions of dark mass as a function of the projected distance from the
clusters' centers are shown in Sect. 4. The comparison with the prediction of 
N-body simulations and analytical models is performed in Sect. 5 to quantify the expected effect of a
sizable population of dark remnants. We summarize and discuss our results in
Sect. 6.

\section{Observational data}
\label{obs_sec}

\begin{table}
 \centering
 \begin{minipage}{140mm}
  \caption{Summary of the adopted radial velocity datasets.}
  \begin{tabular}{@{}lcccr@{}}
  \hline
& & NGC 288 & &\\
  \hline
  Observing & PI & FLAMES  & \# of & \# of\\
  program   &    & setups  & targets & {\it bona-fide}\\
            &    &         &         & targets\\
 \hline
071.D-0131(A) & Moheler & LR1/LR2/LR3 & 20 & 17\\
073.D-0211(A) & Carretta & HR11/HR13 & 117 & 111\\
074.A-0508(A) & Drinkwater & LR2/LR4 & 87 & 57\\
075.D-0043(A) & Carraro & HR9 & 175 & 161\\
087.D-0276(A) & D'Orazi & HR15/HR19 & 78 & 75\\
088.B-0403(A) & Lucatello & HR9 & 88 & 84\\
193.D-0232(D) & Ferraro & HR21 & 174 & 162\\
AAO           & Lane    &      & 1237 & 152\\
\hline
& & NGC 6218 & &\\
\hline
073.D-0211(A) & Carretta & HR11/HR13 & 92 & 82\\
087.D-0276(A) & D'Orazi & HR15 & 76 & 73\\
193.B-0936(I) & Gilmore & HR10 & 108 & 86\\
193.D-0232(B) & Ferraro & HR21 & 385 & 321\\
AAO           & Lane    &      & 2937 & 317\\
\hline

\end{tabular}
\end{minipage}
\end{table}

\begin{figure*}
 \includegraphics[width=18cm]{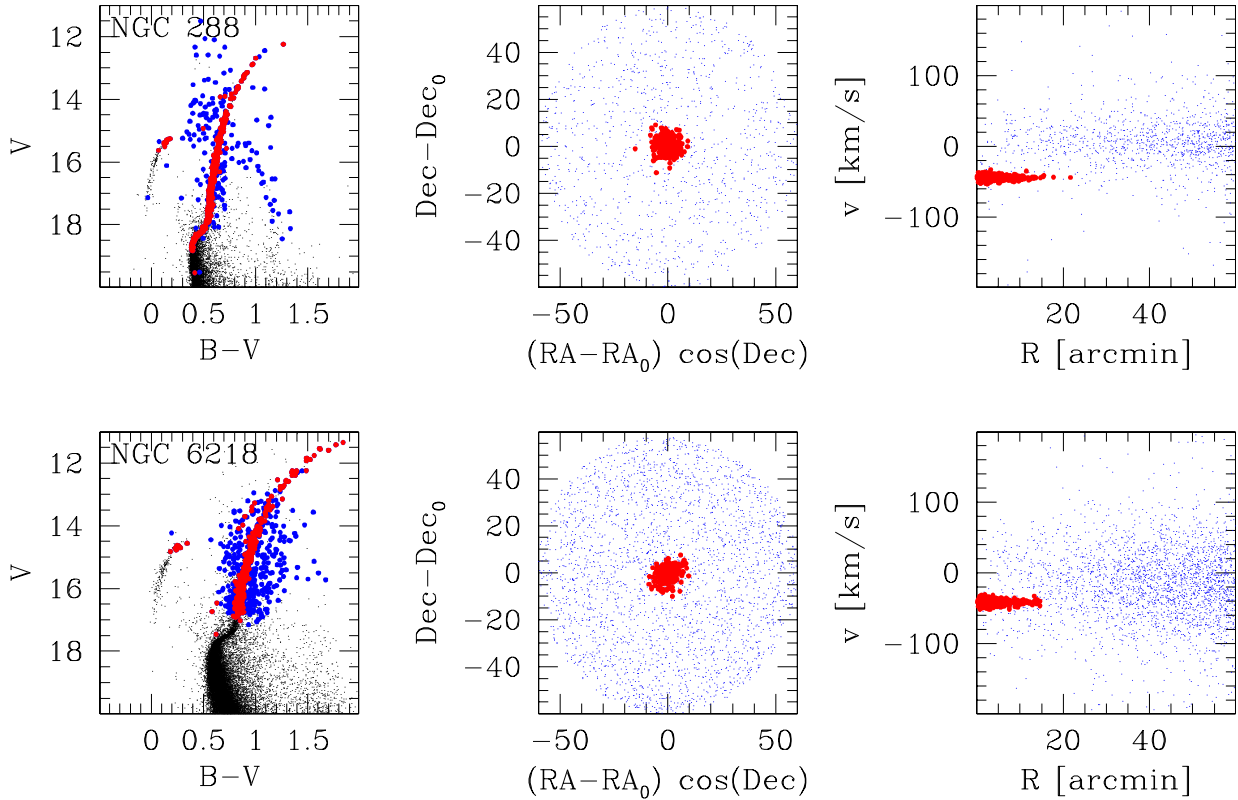}
 \caption{Left panels: WFI CMDs of the two analysed clusters (black dots).
 Middle panels: spatial location of our spectroscopic targets. Right panels:
 radial velocities as a function of the distance from the cluster center.
 In all panels stars with measured radial velocities and bona-fide cluster
 members (see Sect. \ref{met_dyn_sec}) are marked with blue and red dots (dark
 grey and black in the printed version of the paper),
 respectively. Top and bottom panels refer to NGC 288 and NGC 6218,
 respectively.}
\label{cmdmap}
\end{figure*}

The analysis performed here is based on both photometric and spectroscopic
datasets. The main photometric database is constituted by the set of publicly 
available deep photometric catalogs of the 
``globular cluster treasury project" (Sarajedini et al. 2007). It consists of 
high-resolution Hubble Space Telescope (HST) images secured with the Advanced 
Camera for Surveys (ACS) Wide Field Channel through the F606W and F814W filters. 
The field of view of the camera ($202\arcsec \times 202\arcsec$) is centered 
on the cluster's center with a dithering pattern to cover the gap between the 
two chips, allowing a full coverage of the core of both the GCs considered in 
our analysis. This survey provides deep color-magnitude diagrams (CMDs) 
reaching the faint MS of the target clusters down to the 
hydrogen burning limit (at $M_{V}\sim 10.7$) with a signal-to-noise ratio 
S/N$>$10. The results of artificial star experiments are also available to 
allow an accurate estimate of the completeness level and photometric errors. 
A detailed description of the photometric reduction, astrometry, and artificial 
star experiments can be found in Anderson et al. (2008). 
Auxiliary wide-field 
photometric data are needed in our work to determine the radial density profile
of the two analysed clusters across their entire extent. For this purpose we
analysed a set of images
collected with the Wide Field Imager (WFI) mounted on the 2.2m telescope of the
European Southern Observatory (ESO; La Silla, Chile). The WFI covers a total field of view of $34\arcmin \times 33\arcmin$ 
consisting of eight 2048 $\times$ 4096 EEV-CCDs with a pixel size of
$0.\arcsec238$/px. Observations were carried out in 
two photometric nights on October 2001 and June 2002 within the observing programs
68.D-0212(A) (PI: Ferraro) and 69.D-0582(A) (PI: Ortolani) for NGC 288 and NGC
6218, respectively, both aimed at the
characterization of the stellar populations of a sample of Galactic GCs. They
consist of a set of B and V images with a dithering pattern to cover the gaps between 
the CCDs. One shallow and three deep exposures have been secured for each
cluster in both passbands to avoid saturation of the bright red giant branch (RGB) stars and, at the 
same time, sample the faint MS stars with a good signal-to-noise ratio.
The photometric reduction of both datasets has been performed using the
DAOPHOT/ALLFRAME PSF-fitting routine (Stetson 1994).
We performed the source detection on the stack of all images while the 
photometric analysis was performed independently on each image. Only stars 
detected in two out of three long exposures or in the short ones have been 
included in the final catalog. We used the most isolated and brightest stars in 
the field to link the aperture magnitudes to the instrumental 
ones. Instrumental V magnitudes have been transformed into the ACS F606W system 
using a first order linear relation obtained by comparing the stars in common
between the two datasets. The final catalogs have been astrometrically calibrated 
through a cross-correlation with the 2MASS catalog (Skrutskie et al. 2006). 
The astrometric solution has a typical standard deviation of 200 mas.
The final WFI CMDs sample the entire evolved population of both clusters and the MS
down to $V\sim23$ i.e. $\sim$4.5 mag below the turn-off (see Fig. \ref{cmdmap}).
Artificial star
experiments show that the completeness at distance $>1.\arcmin6$ (approximately 
the ACS field of view) is $>90\%$ for $V<20$.

The radial velocity database has been assembled using spectra observed in
different spectroscopic campaign, in particular: 
\begin{itemize}
\item{The bulk of our dataset within the clusters' tidal radii comes from the 
analysis of spectra secured within the large program 193.D-0232 
(PI: Ferraro) aimed at probing the internal dynamics of a sample of 30 Galactic
GCs. Observations have been performed
with the Fibre Large Array Multi-Element Spectrograph (FLAMES; Pasquini et al. 
2002) at the ESO Very Large Telescope used in GIRAFFE mode, using the 
high-resolution ($R\sim 18000$) grating HR21 (8484--9001 $\rm \mathring{A}$). 
Three and seven pointings have been performed in NGC 288 and NGC 6218,
respectively, reaching a 
S/N of 50--300 pixel$^{-1}$, depending on the star 
magnitude.} 
\item{All the FLAMES spectra available at
the ESO archive for the two target clusters have been retrieved and analysed 
using the same technique adopted for the above described dataset. They consist
of surveys of RGB stars observed with different setups and resolutions collected
in the last 13 years. A summary of the databases used in this work is given in
Table 1.}
\item{The above datasets have been complemented with the radial velocity database 
by Lane et al. (2010, 2011)
performed using the multi-fiber spectrograph mounted at the Anglo-Australian
Telescope AAOmega (AAO) which covers a wide area around the target clusters up to a
distance of 1 deg from the clusters' centers, sampling the whole radial extent of the
clusters and the surrounding field population. Spectra were observed with the 1700D and 1500V gratings on 
the red and blue arms, respectively. With this configuration, spectra covering 
the CaII triplet region (8340-8840 $\rm \mathring{A}$) and the interval
containing iron and magnesium 
lines around $\sim5200 \rm \mathring{A}$ were obtained with a resolution of R=10000 and 
R=3700 for the red and blue arms, respectively. A detailed description of the 
reduction procedure and radial velocity estimates can be found in Lane et al. 
(2010, 2011). For the present work, we adopted the radial 
velocities extracted with the RAVE pipeline since they provide a better 
estimate of the radial velocity uncertainty when compared with the available 
high-resolution spectroscopic studies (see Bellazzini et al. 2012).} 
\end{itemize}

Raw FLAMES data have been reduced with the GIRAFFE ESO Base-Line Data Reduction 
Software v2.14.2 (BLDRS) which includes bias subtraction, 
flat-field correction, cosmic rays removal, wavelength calibration and one-dimensional spectra
extraction. The spectra acquired with the fibres
dedicated to sky observations in each exposure have been averaged to obtain a
mean sky spectrum and subtracted from the object spectra by taking 
into account the different fibre transmission. The spectra have been then 
continuum-normalized with fifth-degree Chebyshev polynomials using the IRAF task
{\it continuum}. 
Radial velocities have been derived through Fourier cross-correlation, using 
the {\it fxcor} task in the radial velocity IRAF package. The spectrum of each object 
has been correlated with a high S/N solar spectrum observed with FLAMES with the
same instrumental setup. All spectra have been corrected for heliocentric 
velocity. For the subset of stars that were repeately observed within each 
dataset radial velocities have been averaged and the
corresponding errors on 
the mean of repeated measures have been assigned as their corresponding 
uncertainties. The errors determined from more than 3 measures have been then
compared with the formal errors provided by the cross-correlation algorithm.
We find a scaling factor of $\sigma_{rms}/\sigma_{Xcorr}=0.9$ between these two
estimates which has been used to convert the uncertainties of radial velocities
measured in $<3$ exposures to the same scale of other measures.
Radial velocities from different datasets have been reported to a homogenous
reference system using as reference the AAO sample which
provides a significant ($>20$) overlap with the other datasets. Radial
velocities measured in multiple datasets have been compared to estimate the
occurrence of binarity: as a selection criterion, we flagged as binaries those
stars whose probabilities that the observed scatter is due to statistical
fluctuations (estimated through a $\chi^{2}$ test) is below 1\% (see Lucatello et
al. 2005).
Of the 193 and 295 stars measured in multiple datasets, we found 15 and 20
binaries in NGC 288 and NGC 6218, respectively, in agreement with the binary
fractions measured in these GCs by many authors with different techniques
(Bellazzini et al. 2002; Sollima et al. 2007; Milone et al. 2012; Lucatello et
al. 2015).
Radial velocities of single stars measured in different datasets have been 
then averaged, while binaries have been excluded from the following analysis. 
The effect of the residual contamination from binaries in single-epoch datasets
is discussed in Sect. \ref{bias_sec}.
The final datasets consist of 1586 and 3188 radial velocities in NGC 288 and NGC
6218, respectively, sampling mainly RGB stars
plus some asympthotic giant and red horizontal branch star across the entire 
radial extent of both clusters (see Fig. \ref{cmdmap}). 
The radial velocity uncertainties are generally in the range 
$0.1<\delta_{v}/km~s^{-1}<1.2$ with a mean value of $\langle\delta_{v}\rangle=0.6~km~s^{-1}$.

\section{Method}
\label{met_sec}

The fraction and distribution of the dark mass in the two analysed GCs have been
determined by comparing the mass profile derived by summing the
masses of individual stars estimated from the ACS CMD (the ``luminous mass
profile'') and that estimated through the Jeans analysis of the spectroscopic
dataset (the ``dynamical mass profile''). In the next sections we describe the techniques used to derive
these profiles.

\subsection{Luminous mass profiles}
\label{met_lum_sec}

\begin{figure}
 \includegraphics[width=8.6cm]{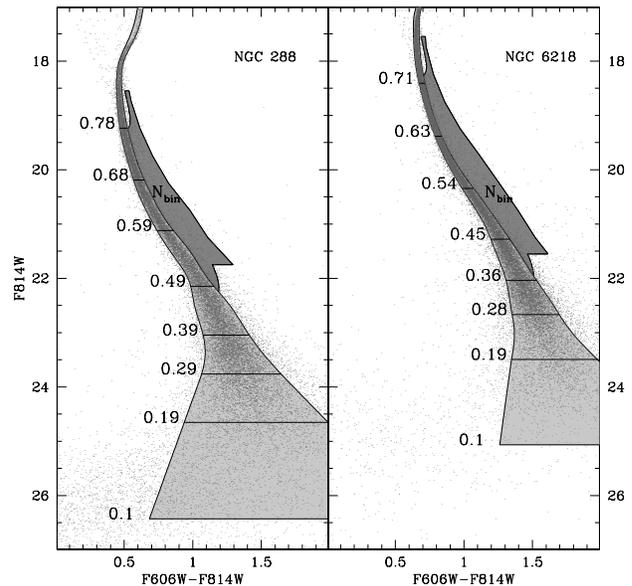}
 \caption{Selection boxes adopted for the population of single stars and 
 binaries (Nbin) of NGC 288 (left panel) and NGC 6218 (right panel). 
 The F814W-(F606W-F814W) color-magnitude 
 diagram is overplotted. The stellar masses at different magnitude levels are
 indicated in both panels.}
\label{cmd}
\end{figure}

The deep CMDs provided by the ``ACS globular cluster treasury project'' sample the
entire unevolved stellar populations of the analysed clusters reaching the
hydrogen burning limit with a completeness $>90\%$. The only objects excluded
from this sample are the dark remnants whose luminosities fall below the detection
limit of ACS observations. GC stars occupy different regions of the CMD
according to their evolutionary stages and masses, so an estimate of their
masses can be made through the comparison with suitable theoretical isochrones. 
We adopted the theoretical isochrones by Dotter et al. (2007) with appropriate ages and
metallicity which have been
converted into the observational color-magnitude plane assuming the distance moduli
and reddening $(m-M)_{0}=14.83$; $E(B-V)=0.01$ for NGC 288 and 
$(m-M)_{0}=13.46$; $E(B-V)=0.19$ for NGC 6218 (Dotter et al. 2010; 
Sollima et al. 2012).

In principle, some contamination from fore/background field stars can
affect star counts at various magnitude levels. However, because of the
relatively high Galactic latitude of the two target clusters ($b<-25^\circ$), 
the density of 
field stars within the cluster core is several orders of magnitude smaller than 
that of the GC population: a comparison with the Galactic model of Robin et al.
(2003) indicates that less than 0.1\% of the stars within the selection boxes 
adopted in our analysis are expected to 
be field contaminants in both the analysed GCs. For
this reason we do not apply any correction for this effect in our analysis.

The effect of unresolved binaries and photometric
errors can have a critical impact by spreading out stars far from their original
location in the CMD,
making difficult to assign them a proper mass. 
The emerging flux from an unresolved binary star is given by the sum of the
fluxes of the two component. According to the mass ratio (q) of the
binary components the star will move in the CMD toward brighter
magnitudes and redder colors with respect to a single star with a mass equal to
the mass of the primary component of the binary.
Because of photometric errors the positions in the CMD of binaries and single
stars partially overlap in the region close to the MS locus where both low-q binaries
and single stars reside. Moreover, the same effect that increases the
binaries magnitudes occurs as a result of chance superpositions between single 
stars, which can occur in the dense central region of GCs.
Although it is not possible to unambiguously distinguish binaries and single
stars across the entire CMD, we adopt a statistical classification of cluster
members. 
For this purpose, the field of view of the ACS data has been divided in
16 annular concentric regions with both width and separation of $0.\arcmin 1$. For
each region a synthetic CMD has been simulated by randomly extracting masses from a 
power-law MF and deriving the corresponding F606W and F814W magnitudes by
interpolating through the adopted isochrone. A population of binaries has been
also simulated by associating to a fraction of stars a secondary
component with a mass randomly extracted from a flat mass-ratios 
distribution (Milone et al. 2012). The fluxes of the two components have been
then summed in both passbands to derive their corresponding magnitudes and color.
For each single and binary system a synthetic star in the same radial range and 
with magnitudes within 0.25 mag has been extracted  
from the library of artificial stars and, if
recovered, its output magnitude and color have been adopted to construct the
synthetic CMD. In this way, the effect of photometric errors, blending and completeness
at the different distances from the cluster center are properly taken into account.
In each radial bin, the MF slope and the binary fraction\footnote{Note that
because of the choice of a fixed mass-ratio distribution for binaries, the
binary fraction and the MF slope do not constitute a degenerate space of 
solutions.} have been tuned to reproduce
the number counts in nine regions of the F814W-(F606W-F814W) CMD defined as
follows: eight F814W magnitude intervals corresponding to 
equal-mass intervals and including all stars with colors 
within three times the photometric error corresponding to their magnitudes, and
a region including the bulk of the binary population with high mass ratios
($q>0.5$). This last region is delimited in magnitudes by the loci of binaries with primary 
star mass $M_{1} = 0.45 M_{\odot}$ (faint boundary) and $M_{1} = 0.75
M_{\odot}$ (bright boundary), and in color by the MS ridge line (blue boundary) 
and the equal-mass binary sequence (red boundary), both redshifted by three 
times the photometric error (see Fig. \ref{cmd}).
A synthetic horizontal branch (HB) has been also simulated for each cluster 
using the tracks by Dotter et al. (2007), tuning the mean mass and mass dispersion 
along the HB to reproduce the observed HB morphology. As a final step, we
associated to each observed star the mass of the closest synthetic object in 
the simulated CMD. We associated to each
observed star a completeness factor ($c_{i}$) defined as the fraction of
recovered artificial stars\footnote{An artificial star has been
considered recovered if its input and output magnitudes differ by less than 2.5
log(2) ($\sim0.75$) mag in both F606W and F814W magnitudes.} with input magnitudes 
in both bands within 0.25 mag and distance from
the cluster center within $0.\arcmin 05$ from those of the
corresponding observed star. The cumulative
radial distribution of luminous mass has been then derived by summing the
completeness-weighted masses of all stars within a given projected distance
\begin{equation}
M_{lum}(R)=\sum_{d_{i}<R} \frac{m_{i}}{c_{i}}
\label{mlum_eq}
\end{equation}

Following the above procedure, the luminous mass enclosed within a projected
distance equal to the extent of the ACS field of view
($1.\arcmin6$) turns out to be $9.3\pm0.9\times10^{3}~M_{\odot}$ and
$1.2\pm0.1\times10^{4}~M_{\odot}$ for NGC 288 and NGC 6218, respectively.
As expected, in both clusters the evolved stellar populations (Subgiant Branch,
RGB, HB and Asympthotic Giant Branch) contribute 
to only a small fraction of the cluster mass (6\% and 8\% in NGC 288 and NGC
6218, respectively).

\subsection{Dynamical mass profiles}
\label{met_dyn_sec}
 
\begin{figure*}
 \includegraphics[width=12cm]{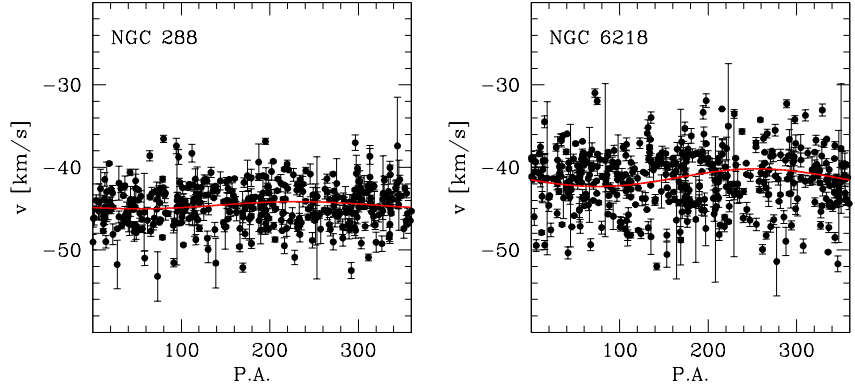}
 \caption{Radial velocity of the {\it bona-fide sample} stars of NGC 288 (left
 panel) and NGC 6218 (right panel) as a function of the position angle. 
 The best-fitting sinusoidal trends are marked with solid lines.}
\label{rot}
\end{figure*}

The dynamical mass profile $M_{dyn}(R)$ has been derived by solving the Jeans
equation in spherical coordinates
\begin{equation}
\frac{1}{\rho_{*}}\frac{d \rho_{*}\sigma_{r,*}^{2}}{d
r}+2\beta\frac{\sigma_{r,*}^{2}}{r}=-\frac{G M(r)}{r^{2}}
\label{jeans_eq}
\end{equation}
where $\rho_{*}$ and $\sigma_{r,*}$ are the 3D mass density and the radial component of
the velocity dispersion of a tracer population at the distance $r$ from the
cluster center, $\beta\equiv 1-\sigma_{t,*}/2\sigma_{r,*}$ is the anisotropy
profile and $G$ is the Newton constant. 
The Jeans equation follows directly from the collisionless Boltzmann
equation and it is based on the concept that any given
population is in hydrostatic equilibrium within the cluster potential. 
This statement holds even if the potential is not completely
generated by the considered population (like in the case of a mass-segregated
stellar system or a DM dominated galaxy). The advantage of the above equation is that it links two observable
quantities (density and velocity dispersion profiles) of an arbitrary sample 
population to a global quantity (the mass profile generating the gravitational 
potential).
Any spherical pressure-supported system at equilibrium must obey to the above 
equation, which allows to derive the dynamical mass profile once the 3D density,
velocity dispersion and anisotropy profiles of a tracer population are
known. 

The observed projected density profiles have been constructed from both ACS and WFI photometric data 
by dividing the number of stars brighter than a threshold magnitude
and contained in circular concentric annuli by the annulus area. The threshold magnitude
has been chosen to include MS and RGB stars with masses $M>0.875
M_{tip}+0.01$, where $M_{tip}$ is the mass at the tip of the RGB extracted from
the best-fit isochrone (see Sect. \ref{met_lum_sec}). This ensures to sample the density profile of stars with similar
masses of those for which radial velocities are available, thus minimizing
variations induced by two-body relaxation while still guaranteeing a good statistics to
sample the outskirts of both clusters. Only stars within 3 times the color
dispersion around the MS/RGB mean ridge line have been used to avoid
contamination from Galactic field stars which could be particularly severe in
the outermost cluster region.
WFI data have been used at distances
$>1.\arcmin6$ while ACS data have been used to sample the
density profile in the innermost region of the clusters where crowding might 
affect completeness in ground based data. The two profiles have been normalized
using stars in common in the outermost bin of ACS data. Star counts have been
corrected for incompleteness using corrections derived from artificial star
experiments. The resulting profiles are consistent with those already published
by other authors (Trager, King \& Djorgovski 1995; Miocchi et al. 2013). 
Angular distances of both photometric and spectroscopic targets
have been converted in physical units adopting the distance moduli and reddening
corrections reported in Sect. \ref{met_lum_sec}.

The analysis of radial velocities has been performed on a {\it bona-fide sample}
defined selecting
stars with {\it i)} radial velocities within 5$\langle\sigma_{LOS}\rangle$ from the mean systemic
velocity, {\it ii)} distance from the clusters' centers smaller than the cutoff
radius apparent in the density profile ($r_{t}$; at $32\arcmin$ and $15\arcmin$ for 
NGC 288 and NGC 6218, respectively) and {\it iii)}
location in the CMD within 5 times the color dispersion about the MS/RGB mean
ridge line. 
The mean systemic
velocity and dispersion ($\bar{v}$ and $\langle\sigma_{LOS}\rangle$) have been derived by maximizing the log-likelihood
\begin{eqnarray}
l&=&\sum_{i} ln \int_{-\infty}^{+\infty}
\frac{exp\left[-\frac{(v'-\bar{v})^{2}}{2\langle\sigma_{LOS}\rangle^{2}}-\frac{(v_{i}-v')^{2}}{2\delta_{i}^{2}}\right]}{2\pi\langle\sigma_{LOS}\rangle\delta_{i}}
dv'\nonumber\\
&=&-\frac{1}{2}\sum_{i}\left(\frac{(v_{i}-\bar{v})^{2}}{\langle\sigma_{LOS}\rangle^{2}+\delta_{i}^{2}}+ln[2\pi(\langle\sigma_{LOS}\rangle^{2}+\delta_{i}^{2})]\right)
\label{sig_eq}
\end{eqnarray}
where $v_{i}$ and $\delta_{i}$ are the velocity of the $i$-th star and its associated 
uncertainty. The derived values of $\bar{v}$ for the two clusters are
$-44.58\pm0.12~
km~s^{-1}$ and $-41.24\pm0.15~km~s^{-1}$ for NGC 288 and NGC 6218, respectively.
By applying the above selection criteria the {\it bona-fide samples} contain 405
and 449 stars for NGC 288 and NGC 6218, respectively.

In principle, systemic rotation affects the dynamical equilibrium of a stellar
system requiring additional terms to be added to eq. \ref{jeans_eq}.
In Fig. \ref{rot} the radial velocities of the {\it bona-fide sample} as a function of the
position angle are shown for the two analysed clusters. The distributions 
appear homogeneous with no apparent trends. A fit with sinusoidal curves 
indicate maximum rotation amplitudes of $v_{rot}~sin~i=0.43\pm0.48~km~s^{-1}$ for NGC 288 
and $v_{rot}~sin~i=1.07\pm0.72~km~s^{-1}$ for NGC
6218, consistent in both cases with no significant 
rotation along the line of sight (see also
Bellazzini et al. 2012). For this reason rotation
terms in the Jeans equation have been neglected for both clusters.

In the literature, many groups inverted the Jeans equation adopting 
smoothed density and velocity dispersion profiles (Gebhardt \& Fischer 1995; L{\"u}tzgendorf et al. 2013; Ibata et al.
2013; Watkins et al. 2015a). There are however two drawbacks in this approach: {\it i)}
according to eq. \ref{jeans_eq}, the derived mass profile depends on the
derivatives of the density and velocity dispersion profiles. So, any fluctuation
due to Poisson noise in observational data produces an unphysical
behaviour of the derived mass profile; {\it ii)} a given $M(r)$ profile can
be a solution of the Jeans equation even if it provides a non-physically 
meaningful representation of the system i.e. corresponding to a distribution 
function which is negative somewhere in the energy domain.
To overcome the above issues we fit both the projected density and 
LOS velocity dispersion profiles of our tracer population with parametric 
analytical functions, by exploring only the portion of the parameter space
providing physically meaningful solutions (i.e. those corresponding
to positive distribution functions; see appendix A).

A particularly convenient choice is to model the derived 3D density and
square velocity dispersion profiles with a sum of Gaussian functions with
different dispersions (Cappellari et al. 2002)
\begin{eqnarray}
\rho_{*}&=&\sum_{j} \mu_{j} exp\left(-\frac{r^{2}}{2 s_{j}^{2}}\right)\nonumber\\
\sigma_{r,*}^{2}&=&\sum_{j} k_{j} exp\left(-\frac{r^{2}}{2 s_{j}^{2}}\right)
\label{mge_eq}
\end{eqnarray}
so that their radial derivatives can always be expressed as sums of
functions with the same coefficients
\begin{eqnarray}
\frac{d^{n}\rho_{*}}{d r^{n}}&=&\sum_{j} \mu_{j}\frac{d^{n}}{d r^{n}}~exp\left(-\frac{r^{2}}{2
s_{j}^{2}}\right)\nonumber\\
\frac{d^{n}\sigma_{r,*}^{2}}{d r^{n}}&=&\sum_{j}
k_{j}\frac{d^{n}}{d r^{n}}~exp\left(-\frac{r^{2}}{2 s_{j}^{2}}\right)
\label{mge3_eq}
\end{eqnarray}
The 3D mass and global density profiles can be then derived from eq. 
\ref{jeans_eq}, \ref{mge_eq} and \ref{mge3_eq} as
\begin{eqnarray}
M(r)&=&-\frac{\sigma_{r,*}^{2}r^{2}}{G}\left(\frac{1}{\rho_{*}}\frac{d\rho_{*}}{dr}+\frac{1}{\sigma_{r,*}^{2}}\frac{d\sigma_{r,*}^{2}}{dr}+\frac{2\beta}{r}\right)\nonumber\\
\rho(r)&=&\frac{1}{4\pi r^{2}}\frac{dM}{dr}
\label{rho_eq}
\end{eqnarray}
In the same way, high order derivatives of density can be also analitically
calculated from the above equation.
The advantage of this choice is twofold: first, once the coefficients $\mu_{j}$
and $k_{j}$ are determined,
the corresponding projected density and LOS velocity dispersion of the tracer
population can be directly computed as
\begin{eqnarray}
\Gamma(R)&=&\sum_{j} \mu_{j} \int_{R}^{+\infty} \frac{r~e^{\frac{r^{2}}{2
s_{j}^{2}}}}{\sqrt{r^{2}-R^{2}}} dr\nonumber\\
\sigma_{LOS}^{2}(R)&=&\frac{1}{\Gamma(R)} \sum_{j} k_{j}\int_{R}^{+\infty} \rho_{*}
e^{\frac{r^{2}}{2 s_{j}^{2}}} \frac{r^{2}-\beta R^{2}}{r \sqrt{r^{2}-R^{2}}}
dr
\label{mge2_eq}
\end{eqnarray}
respectively. Second, the derived 3D density and velocity dispersion profiles are already smooth
functions of radius and their integrals and derivatives can be 
quickly computed analitically.
The Gaussian widths $s_{j}$ (the same for both the density and velocity
dispersion) have been chosen to increase in logarithmic steps of 0.1 from 1 pc
to the cut radius.
The $\mu_{j}$ coefficients were first derived through a $\chi^{2}$ minimization
between the observed projected density profile of tracers and that derived 
through eq. \ref{mge_eq} and \ref{mge2_eq}.
Then, the space of $k_{j}$ coefficients has been sampled with a
Metropolis-Hasting Markov Chain Monte Carlo (MCMC) technique to reproduce the
distribution of velocities in the $v-R$ plane. 
Note that residual contamination from fore/background field stars can be present in the
{\it bona-fide sample}, in particular in the outer portions of the clusters where the
density of the tracer population falls below the density of field contaminants.
To account for this effect we modelled the population of field stars with a
Gaussian whose mean velocity and dispersion ($\bar{v_{f}}$ and $\sigma_{f}$) 
have been derived by applying the maximum-likelihood
technique (eq. \ref{sig_eq}) to stars beyond the cutoff radius. Eq. \ref{sig_eq} has been then
modified to account for the field contamination in the following way
\begin{equation}
l=\sum_{i} ln
\left(\frac{exp\left[-\frac{(v_{i}-\bar{v})^{2}}{2(\sigma_{LOS}^{2}+\delta_{i}^{2})}\right]}
{\sqrt{2\pi(\sigma_{LOS}^{2}+\delta_{i}^{2}})}+
\frac{\Gamma_{f}exp\left[-\frac{(v_{i}-\bar{v_{f}})^{2}}{2(\sigma_{f}^{2}+\delta_{i}^{2})}\right]}
{\Gamma(d_{i})\sqrt{2\pi(\sigma_{f}^{2}+\delta_{i}^{2}})} \right)
\label{sig_eq2}
\end{equation}
where $\Gamma_{f}$ is the projected density of field
stars. Because of the relatively small area covered by our 
observations we adopted a constant value of $\Gamma_{f}$ which has been calibrated
to reproduce the fraction of stars within and outside
5$\langle\sigma_{LOS}\rangle$ from the 
mean systemic velocity and in the radial range between $0.5<R/r_{t}<1$. 
For each MCMC step a given combination of
$k_{j}$ parameters is analysed and both the distribution
function $f(Q)$ (using the Eddington 1916 formula; see Appendix) and the 
log-likelihood (from eqs. \ref{mge_eq}, \ref{mge2_eq} and \ref{sig_eq2}) are calculated. 
At the end of the MCMC cycle all combinations of $k_{j}$ producing a
negative $f(Q)$ in any region of the energy domain are rejected and the best-fit
values of $k_{j}$ are chosen as those where the majority of realizations are
placed.
The cumulative projected mass profile has been then calculated using eqs. 
\ref{mge_eq}, \ref{mge3_eq}, \ref{rho_eq} and
$$M_{dyn}(R)=4\pi\int_{0}^{R} R' \int_{R'}^{+\infty}
\frac{\rho~r}{\sqrt{r^{2}-R'^{2}}} dr~dR'$$
Uncertainties in the derived profile have been calculated in a Monte Carlo
fashion: for each observed star in the {\it bona-fide sample} a synthetic velocity has been
extracted from a Gaussian with dispersion equal to the convolution of the
best-fit LOS velocity dispersion at the star distance and
its observational uncertainty. The mass profile has been then calculated on the
so-obtained synthetic sample using the same procedure adopted for real data. One
hundred synthetic samples have been extracted and the r.m.s of the derived
masses at different radii have been adopted as uncertainties.
 
\begin{figure*}
 \includegraphics[width=12cm]{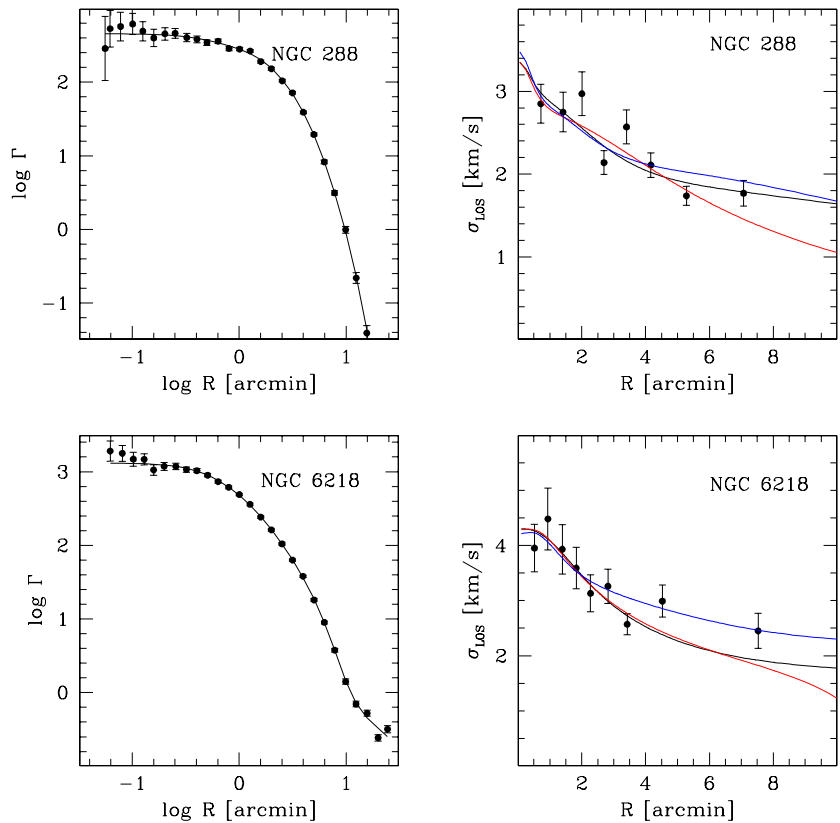}
 \caption{Projected density profiles (left panels) and LOS velocity dispersion 
 profiles (right panels) of NGC 288 (top panels) and NGC 6218 (bottom panels). 
 The best-fit solution obtained by assuming isotropy, radial and tangential anisotropy are shown as 
 black, red and blue lines (solid, dashed and dot-dashed lines in the printed
 version of the paper), respectively. The LOS velocity dispersion profiles have
 been calculated only for illustrative purposes by binning the velocities of the {\it bona-fide sample} in
 groups of 50 stars.}
\label{fit}
\end{figure*}

\begin{figure*}
 \includegraphics[width=12cm]{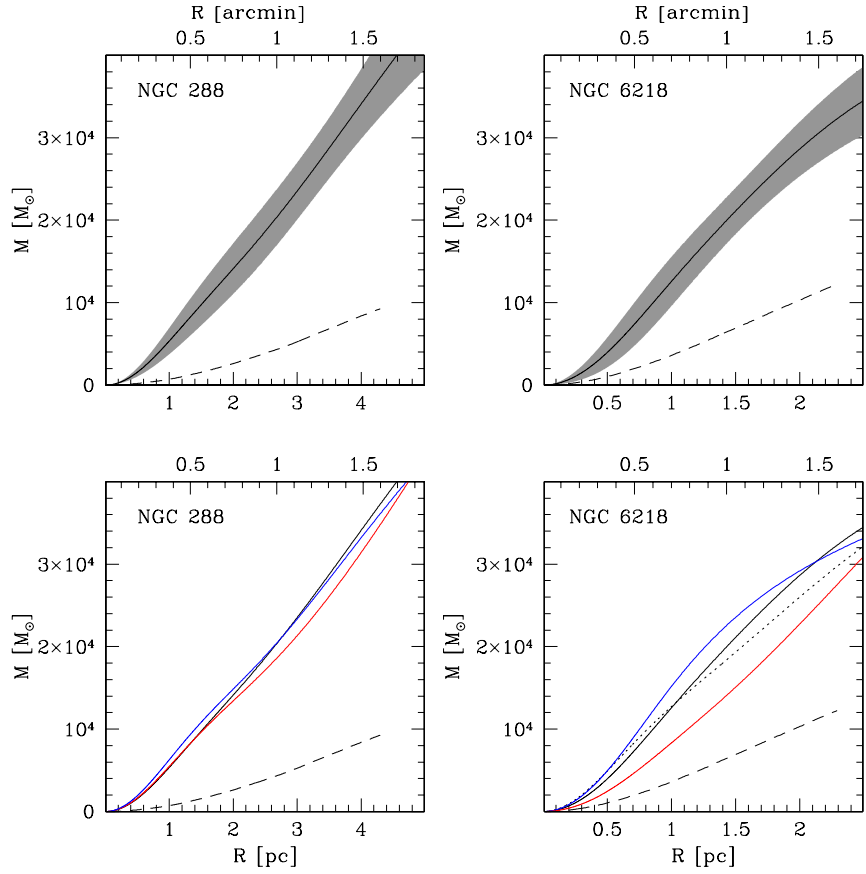}
 \caption{Top panels: cumulative 
 luminous (thick dashed lines) and dynamical (solid lines) mass 
 profiles derived for NGC 288 (left panels) and NGC 6218 (right panels).
 The 1$\sigma$ uncertainties about the isotropic model best-fit are
 marked as grey area. Bottom panels: same as top panels where anisotropic fits
 are shown. The
 color code for anisotropic models is the same adopted in Fig. \ref{fit}.
 The dotted line in the right panel indicates the dynamical mass profile
 calculated using only stars with repeated exposures. In all panels the distance from the cluster center is
 reported in both angular (top scale) and physical units (bottom scale).}
\label{mprof}
\end{figure*}

\begin{figure*}
 \includegraphics[width=12cm]{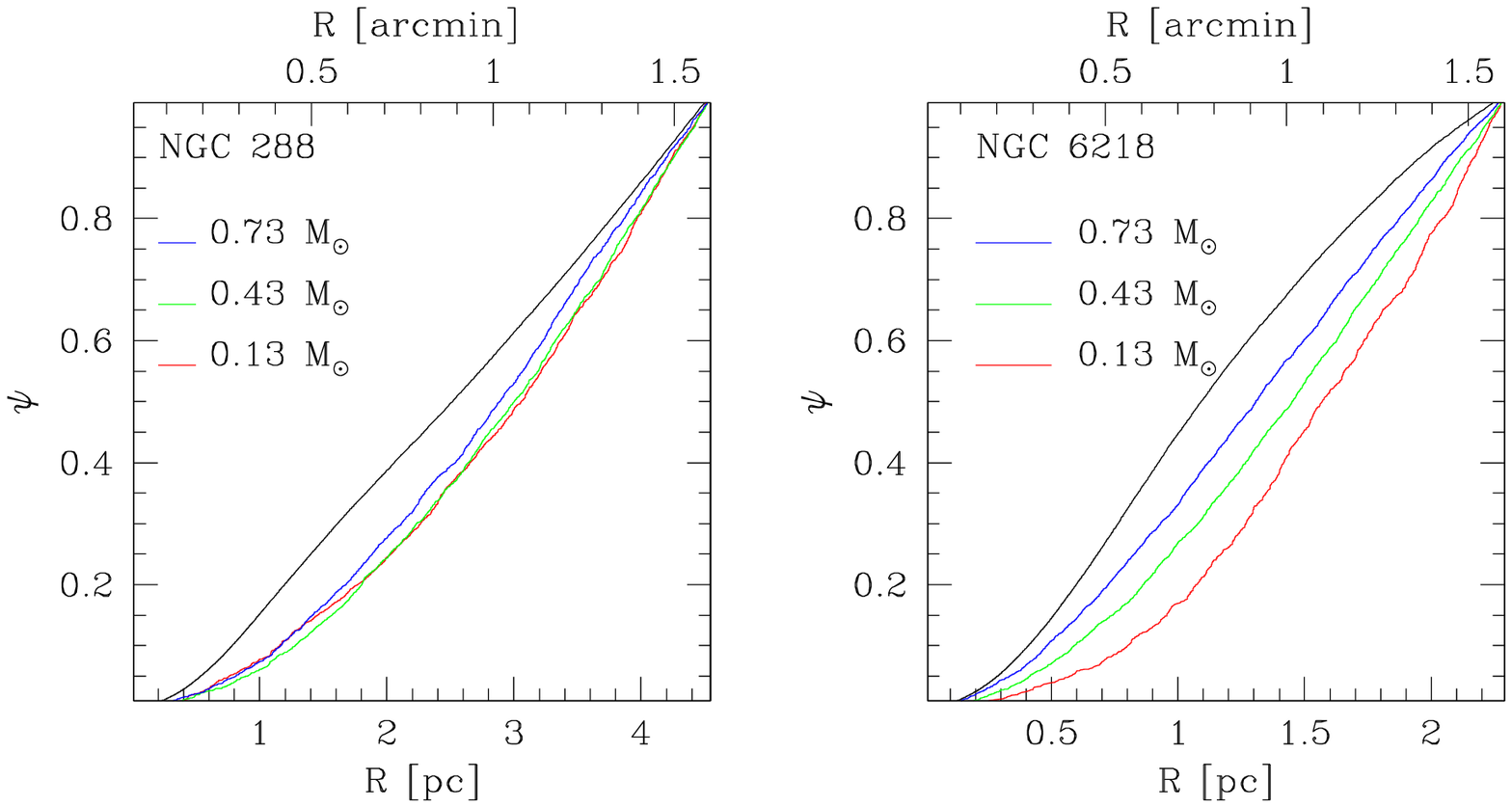}
 \caption{Comparison between the cumulative profile of dark mass (black lines)
 with those of stars in different mass bins in NGC 288 (left panel) and NGC 6218
 (right panel). All profiles are normalized at a distance corresponding to the
 ACS field of view ($1.\arcmin6$).}
\label{cumu}
\end{figure*}

Regarding the anisotropy profile, we considered three cases: isotropic
($\beta=0$ across the entire cluster extent), a radially anisotropic 
Osipkov-Merrit profidle (Osipkov 1979; Merritt 1985) and a tangentially anisotropic profile.
In the considered anisotropic models the $\beta$ profile is 
$$\beta(r)=\pm\frac{\tilde{r}^{2}}{1+\tilde{r}^{2}}$$
with $\tilde{r}=r/r_{a}$, where $r_{a}$ is a characteristic radius which sets
the boundary where orbits become significantly radially/tangentially biased 
according to the sign of the right-hand side in the above equation. The choice of these
profiles has been made to allow to apply the Eddington (1916) formula even in
case of $\beta\neq 0$. The value of $r_{a}$ has been set to the
minimum value satisfying the criterion for stability against radial-orbit instability 
by Nipoti, Londrillo \& Ciotti (2002): $\xi_{half} = 2 T_{r}/T_{t}<1.5$ where $\xi$ is 
the Fridman-Polyachenko-Shukhman parameter (Fridman \& Polyachenko 1984), and
$T_{r}$ and $T_{t}$ are the radial and tangential component of the kinetic energy tensor
calculated within the half-mass radius. In this way we expect to bracket the
entire range of physically meaningful and stable models able to reproduce the
structure and kinematics of these two GCs.

The best fits to the projected density and LOS velocity dispersion profiles of the
two analysed clusters are shown in Fig. \ref{fit}.

\section{Results}
\label{res_sec}

The luminous and dynamical mass profiles of the two analysed GCs within the ACS
field of view are shown in Fig. \ref{mprof}. It is apparent that in both clusters dynamical
masses are significantly larger than luminous ones across the entire 
surveyed area. In particular, at a projected distance of $1.\arcmin6$ (the
extent of the ACS field of view) the 
enclosed dynamical masses are $\sim3.7\times10^{4} M_{\odot}$ and $\sim3.2\times10^{4} M_{\odot}$
in NGC 288 and NGC 6218, respectively i.e. $\sim$3 times larger than the
corresponding luminous masses ($\sim9.3\times10^{3} M_{\odot}$ and $\sim1.2\times10^{4} M_{\odot}$).
Although the uncertainties on the dynamical mass estimate are quite large
($\sigma_{M}\sim3.8\times10^{3} M_{\odot}$) the above discrepancy stands at a
significance of $\sim6~\sigma$.
The estimated masses translate into fractions of dark mass within this distance 
of 75$\pm$12\% and 62.5$\pm$9.6\% in NGC 288 and NGC 6218, respectively.

From the mass profiles derived above it is possible to determine the
distribution of dark mass within the area surveyed by our analysis by simply subtracting at each
radius the enclosed luminous mass to the dynamical one. The resulting cumulative
distribution of dark mass normalized at $1.\arcmin6$ is compared to those of stars in different mass bins
in Fig. \ref{cumu}. For this purpose we applied eq. \ref{mlum_eq} to stars selected 
from the photometric database 
with mass (derived as described in Sect. \ref{met_lum_sec}) comprised 
in three bins of 0.1 $M_{\odot}$ width.
The typical signature of mass segregation is apparent in Fig. \ref{cumu}, with
the most massive stars being more concentrated than less massive ones.
It is apparent that for both clusters the dark mass is more concentrated than
any other mass group. Unfortunately, because of the small number of stars in the
innermost pc (i.e. the smallest dispersion of the gaussian functions adopted to fit the
observed profiles; $s_{1}=1~pc$), we are not able to resolve the shape of the the dynamical mass
density profile within this region (see Sect. \ref{darkremn_sec}).

In the following subsections we will analyse the possible sources of systematics
potentially affecting the above result.

\subsection{Possible observational biases}
\label{bias_sec}

\begin{figure}
 \includegraphics[width=8.6cm]{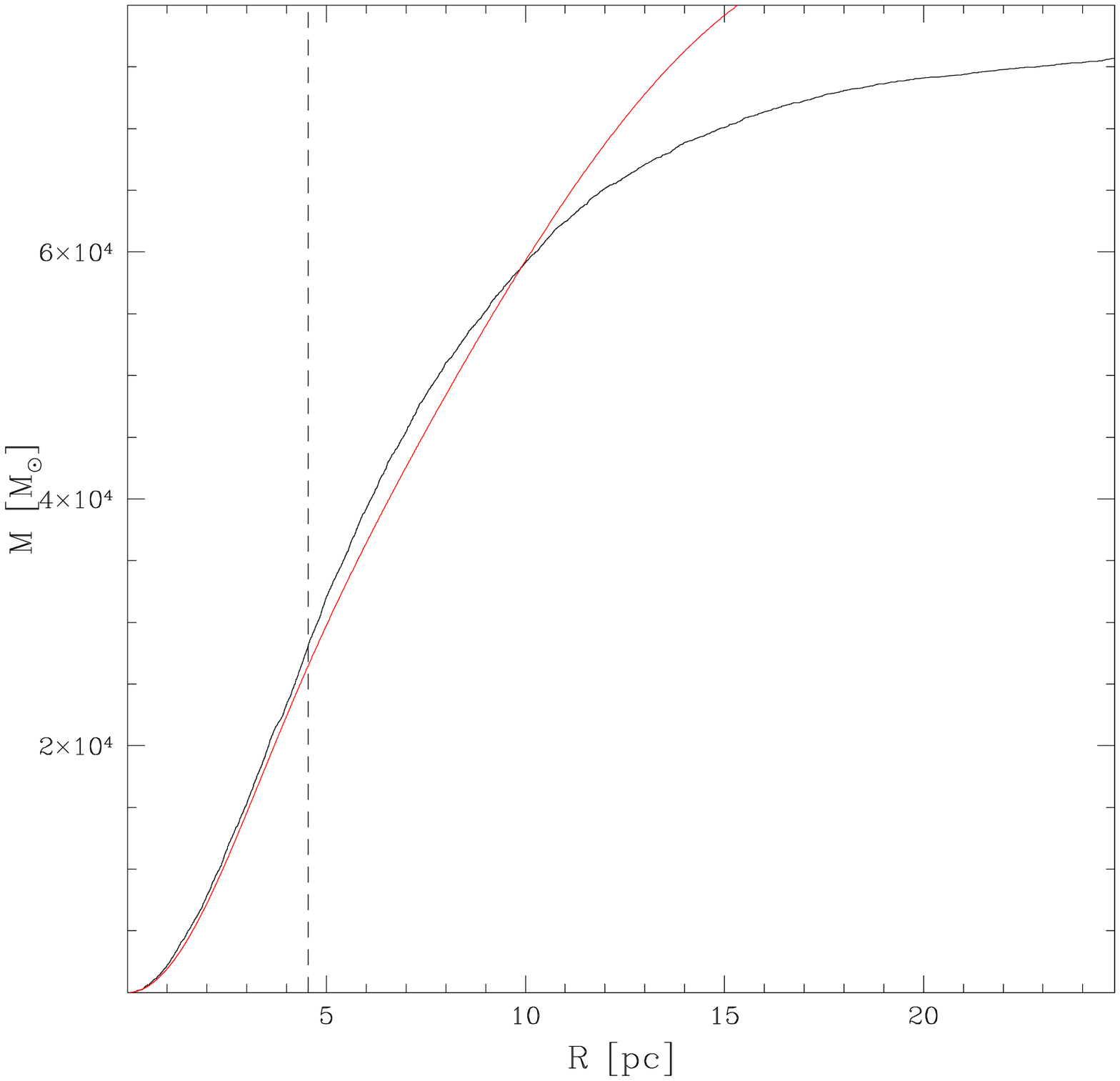}
 \caption{Comparison between the cumulative projected mass profile of 
 the N-body simulation of NGC 288 (black line) and that derived from the Jeans
 analysis (red line; grey in the printed version of the paper). The vertical
 dashed line indicates the radial extent of the present work.}
\label{nbody}
\end{figure}

As already discussed in Sollima et al. (2012), the systematics possibly
affecting the determination of the luminous
mass (like uncertainties on distance, reddening, age, mass-luminosity relation
and photometric completeness) have negligible impact unless variations many
orders of magnitudes larger than the formal uncertainties are present.
In this regard, the dark mass fraction ($f_{dark}$) is less sensitive to the
uncertainties in the above assumptions than the M/L ratio.
For instance, an incorrect choice of distance (the most uncertain among the
above mentioned parameters) would affect both the
luminous and dynamical masses in the same direction. Indeed, the luminous mass
estimated according to eq. \ref{mlum_eq} depends on the masses of individual
stars estimated from their luminosities. Considering that luminosities scale
with distance as $L\propto d^{2}$ and that at low
stellar masses $L\propto M^{3}$, this translates into $M_{lum}\propto d^{2/3}$.
On the other hand, the cluster physical radius is proportional to the adopted 
distance. Following the virial theorem, the same observed velocity
dispersion is reproduced by systems with a constant ratio $M_{dyn}/r$, leading
to $M_{dyn}\propto d$. So, the luminous-to-dynamical mass ratio has only a 
small dependence
on distance ($M_{lum}/M_{dyn}\propto d^{-1/3}$).
Finally, applying the error propagation law one finds
$$\sigma(f_{dark})=\frac{(1-f_{dark})}{3}\frac{\sigma(d)}{d}$$
Considering that GC distances have a typical uncertainty of 5\% and that 
$f_{dark}\sim65\%$ this translates into an error of only $\sim0.6\%$ in the 
derived dark mass fraction.

A potentially significant effect can instead be due to the physical processes
altering the cluster kinematics.
According to eq. \ref{jeans_eq} anisotropy plays a role in shaping the 
dynamical mass profile. 
However, as shown in Fig. \ref{mprof}. the derived dynamical mass profile
appears to be almost independent on the degree of anisotropy.
This is a consequence of the behaviour of the adopted anisotropy profiles which
are almost isotropic within the core and deviate from isotropy only outside
the area covered by our data. Although the functional form of the anisotropy
profile is arbitrary, recent studies based on HST proper motion analyses in a
wide sample of Galactic GCs, including NGC 288, seem to confirm this trend (Watkins et al. 2015b).

Another source of uncertainty is given by unresolved binaries. Indeed, 
in a binary system the relative projected velocity of the primary 
component is added to the motion of the center of mass, introducing an 
additional spread in the velocity distribution of the whole population.
This effect is maximized in the innermost cluster region where
binaries are expected to sink because of their large systemic masses.
As explained in Sect. \ref{obs_sec}, binaries showing radial velocity
variations in multiple datasets have been excluded from the sample. 
Of course, the binary selection criterion fails for binaries with radial 
variation amplitude comparable to the observational error. However, in this case
the spread produced by undetected binaries would be $<0.3~km~s^{-1}$ producing a
negligible effect on the global velocity dispersion in the central region of
these clusters.
On the other
hand, about half of the stars in the {\it bona-fide sample} are present only in 
a single dataset. For these stars also binaries with large radial velocity
variations cannot be detected. To test the effect of this contamination we
performed the same analysis using only the binary-cleaned sample of stars with 
multiple measures for NGC 6218.
The resulting mass profile appears to be almost
indistinguishable from that derived using the entire {\it bona-fide sample} (see
Fig. \ref{mprof}). So, we conclude that binaries alone cannot produce the
observed discrepancy between luminous and dynamical mass. 
Unfortunately, only few stars with
multiple observations are located in the innermost region of NGC 288 thus preventing the
application of the same analysis in this cluster.

\begin{figure*}
 \includegraphics[width=12cm]{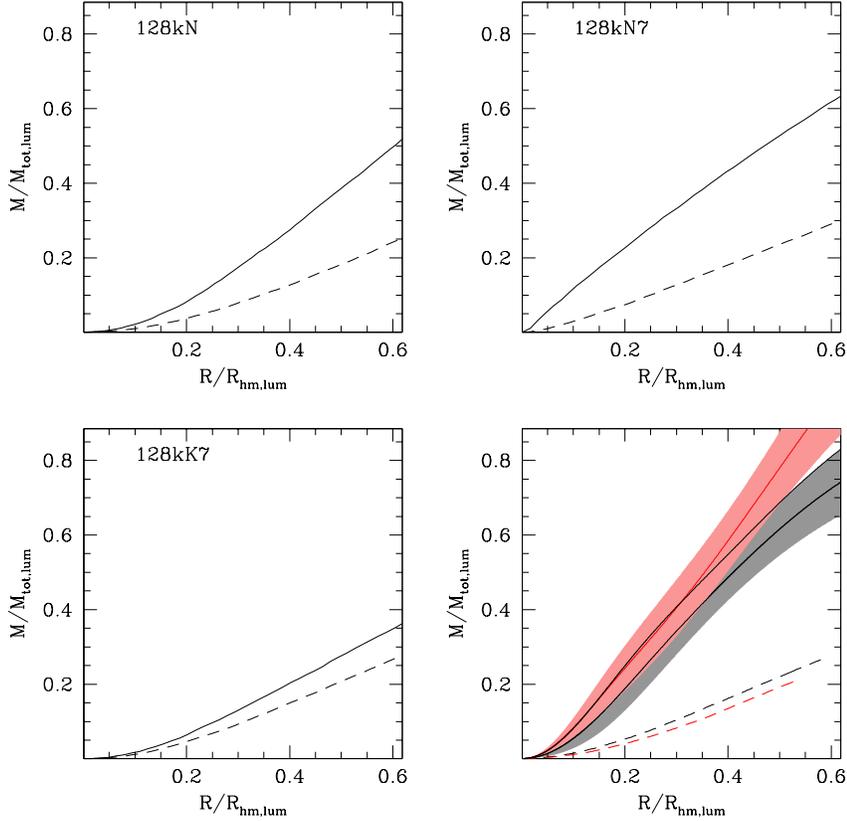}
 \caption{Comparison between the total cumulative mass profile (solid lines)
 with that of luminous stars (dashed lines) in simulations 128kN (top-left
 panel), 128kN7 (top-right panel), 128kK7 (bottom-left panel). The luminous
 and dynamical mass profiles 
 estimated for NGC 288 (red lines, light grey in the printed version of the
 paper) and NGC 6218 (black lines) are shown in the bottom-right panel with
 dashed and solid lines, respectively. The shaded area around solid lines
 indicate the 1$\sigma$ uncertainty on the dynamical mass profiles. The range in projected
 distance is the same of Fig. \ref{mprof}}
\label{nbdark}
\end{figure*}

Finally, the heating due to the tidal interaction of the clusters with the 
Milky Way can also affect the dynamical mass estimate.
Indeed, the periodic shocks occurring every disk crossing and pericentric 
passage transfer kinetic energy to cluster stars, inflating their velocity 
dispersion. While this 
effect is particularly important in the outermost regions (see e.g.
Weinberg 1994; Heggie \& Hut 2003 and references therein), 
K{\"u}pper et al. (2010) have demonstrated, by means of N-body simulations, 
that a population of potential escapers may have significant effects on the 
kinematical properties of star clusters, even at intermediate radii.
To test this hypothesis, we performed the Jeans analysis on the outcome of the 
N-body simulation run by Sollima et al. (2012) with the orbital and structural 
characteristics of NGC 288. 
Particles positions and velocities from the last snapshot of
the simulation have been projected into 
the x-y plane and a subsample of 450 particles has been randomly extracted.
Random velocity offsets extracted from a Gaussian distribution with dispersion equal to
$0.5~km~s^{-1}$
have been added to mimic the effect of observational errors. The derived mass
profile is compared with the actual mass distribution of the simulation in Fig.
\ref{nbody}. It can be seen that the mass distribution is well recovered within
10\% at distances $R<10$ pc (i.e. $\sim$2.5 times larger than the area covered by
our analysis), while at larger distances tidal heating lead to an
improper estimate of the enclosed mass. In conclusion, while tidal heating can
produce significant effects in the LOS velocity dispersion (see Fig. 8 of
Sollima et al. 2012) and in the estimate of the enclosed mass outside the
half-mass radius of NGC 288, it cannot produce the overabundance of
dynamical mass estimated in this paper.

\section{Effect of dark remnants}
\label{darkremn_sec}

\begin{figure*}
 \includegraphics[width=12cm]{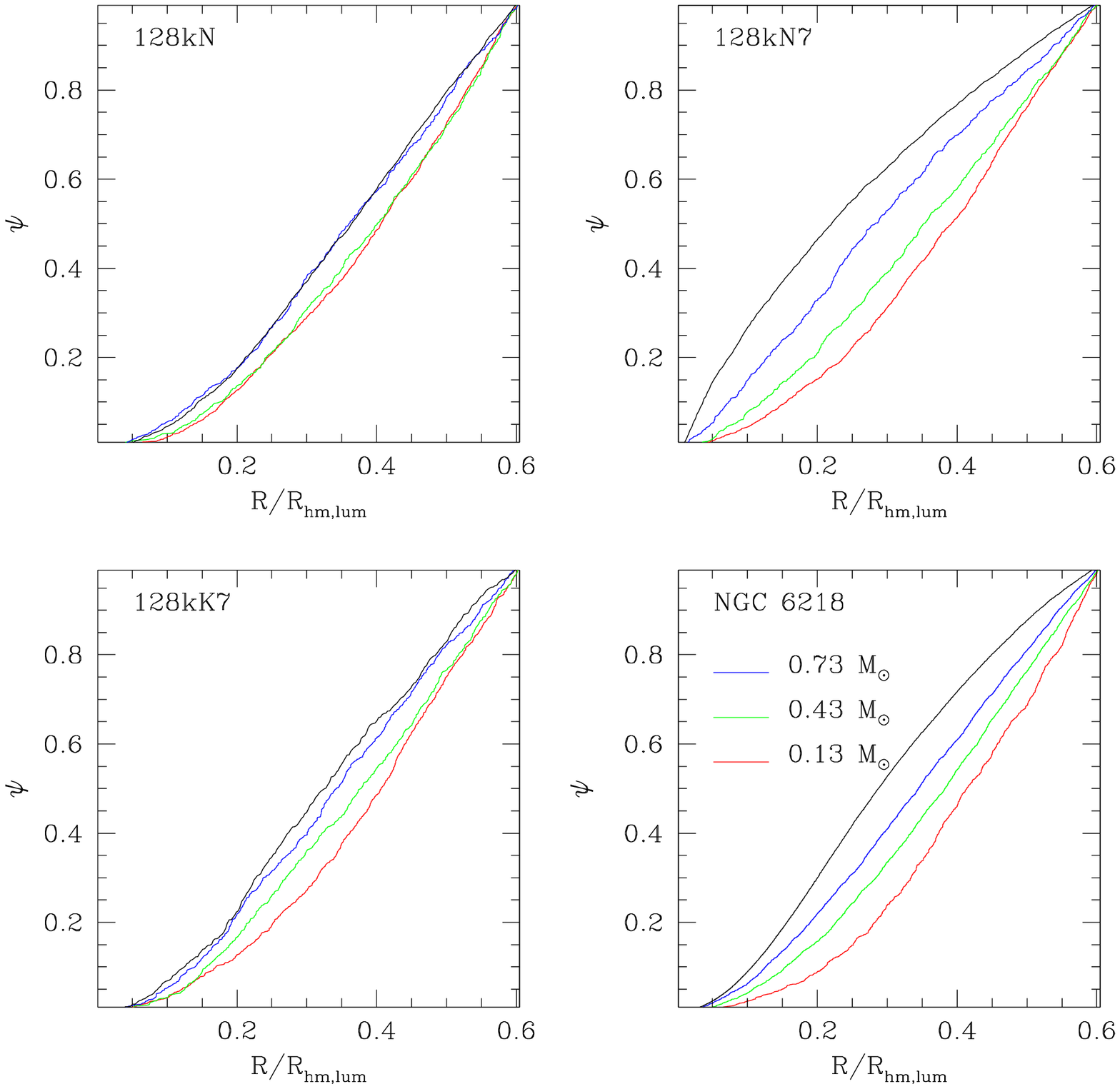}
 \caption{Comparison between the cumulative profile of dark mass (black lines)
 with those of stars in different mass bins in simulations 128kN (top-left
 panel), 128kN7 (top-right panel), 128kK7 (bottom-left panel). The profiles 
 estimated for  NGC 6218 are shown in the bottom-right panel. All profiles are
 normalized at a projected distance of $R/R_{hm,lum}=0.6$. The range in 
 projected distance is the same of Fig. \ref{mprof}}
\label{nbdistr}
\end{figure*}

Since we have ruled out any significant effect of the possible sources of bias 
in our estimate, the observed
discrepancy between luminous and dark mass needs to be interpreted on the basis
of physical grounds.
As discussed in Sect. \ref{intro_sec}, a natural reservoir of non-luminous mass 
is constituted by dark remnants. In this section we try to test whether the
observed overabundance of mass can be entirely due to these objects.

For this purpose we analysed three N-body simulations selected from the set
of Contenta et al. (2015) and compared their fraction of remnants with that
measured in NGC 6218. Simulations adopt 131,072 particles
extracted from a Kroupa (2001) mass function in the mass range
$0.1<M/M_{\odot}<15$ and distributed following a King 
(1966) profile with central
adimensional potential $W_{0}$=5 and 7, and a tidal radius equal to the 
Jacobi radius imposed by the external field. The cluster moves on a
circular orbit within a logarithmic potential with circular velocity
$v_{circ}=220~km~s^{-1}$ at a distance of 8.5 kpc from the
center. Simulations were run with NBODY6 (Nitadori \& Aarseth 2012) including 
the effect of stellar evolution with and without the inclusion of natal kicks to
NSs. From the entire set of simulations discussed in Contenta et al.
(2015) we extracted the snapshot at t=13 Gyr (the age of NGC 6218; Marin-Franch
et al. 2009) from simulation 128kN, 12k8N7 and 128kK7. Indeed, the fraction of
remnants is highly sensitive to the age of the cluster (see Sect.
\ref{intro_sec}), and these are the only simulations of the considered dataset
that survived to dissolution after such time. Simulation 128kN differs from
simulations 128kN7 and 128kK7 because of its lower initial concentration
($W_{0}=5$ instead of 7), while simulation 128kK7 assumes natal kicks for NSs, 
at odds with the other simulations.
It is worth stressing that these simulations after 13 Gyr have very 
different structure, mass, and MF with respect to those of the observed clusters. In
particular, their total masses are $<3\times10^{4} M_{\odot}$ (i.e. a factor
$\sim$ 3 smaller than those of the analysed clusters), and very
different projected half-mass radii and MFs. Moreover, the simplified orbits
followed by simulations could make them subject to a different tidal stress. 
Under these conditions, their
relaxation timescales, efficiencies of dark remnant retention and mass loss
rates are likely different from those of the real clusters. On the other hand, in
absence of specifically calibrated simulations,
this comparison provides a first-order guess of the typical remnant mass
fraction and their radial distribution in these stellar systems.

Particle positions have been projected
on the x-y plane and the actual cumulative mass of unevolved stars and
remnants have been calculated. Of course, the mass, size and
structural properties of the analysed snapshots are substantially different from
those of NGC 6218, so to compare the profiles derived from simulations
with observations we normalized masses to the total luminous mass and projected 
distances to the half-mass projected radius of luminous stars ($R_{hm,lum}$). 
These quantities
can be in fact easily calculated in simulations and can be estimated for NGC
6218 from the best-fit multimass model by Sollima et al. (2012).
In Fig. \ref{nbdark} the luminous and total mass profiles of the three 
analysed snapshots are compared with those observed in the two analysed
clusters. Note that while
in simulation 128kK7 dark remnants constitute only a small fraction ($\sim25\%$) 
of the total mass within $0.6~R_{hm,lum}$, in both simulations 128kN and
128kN7 dark remnants represent $\sim$52\% of the
total mass within the same radius, only slightly
smaller (within 2$\sigma$) than those estimated in NGC 288 ($75\pm12\%$) 
and NGC 6218 ($62.5\pm9.7\%$). It is important to
emphasize that even in simulations where natal kicks are not included the large
majority of mass in remnants is constituted of WDs ($\sim87\%$). So, the large
fraction of dark remnants in these simulations is not due to an increased
retention of NSs but likely to the role played by the presence of NSs in the 
initial phase of dynamical evolution of the
cluster (see Fig. 2 and 4 of Contenta et al. 2015).
As shown in a number of studies (see e.g. Vesperini \& Heggie 1997; 
Baumgardt \& Makino 2003), a significant loss of stars is necessary to explain a large fraction
of remnants: the passive evolution of a Kroupa (2001) initial mass
function in the range $0.1<M/M_{\odot}<120$ (neglecting dynamically induced
losses of stars and NS/BH ejection\footnote{This fraction reduces to 24\% if
a complete ejection of NS and BH is assumed. This can be considered a strong lower
limit since all the neglected dynamical processes (retention of NS/BH,
central segregation of remnants, preferential loss of low-mass MS stars, etc.) 
lead to an increase of this fraction.}) 
would produce a remnant fraction of only $\sim28\%$ after 13 Gyr which can only 
slightly increase as a result of mass segregation of NS, BH and WD progenitors. 
In this regard, the simulations 128kN, 128kN7 and 128kK7 considered here after 
13 Gyr lost 77\%, 70\% and 86\% of their initial mass.
In this context, estimates of the evaporation rates of these two GCs have been
obtained by Kruijssen \& Mieske (2009) and Webb \& Leigh (2015), providing
indications of a significant mass loss ($>80\%$) in both of them.

In Fig. \ref{nbdistr} the cumulative distribution of dark mass normalized 
at $R/R_{hm,lum}=0.6$ (corresponding to a projected distance of $\sim1.\arcmin6$
in both the GCs analysed in this work) is compared with those of different mass groups in all
the considered N-body simulations and in NGC 6218, as an example.
Qualitatively similar considerations
can be made from the comparison with NGC 288 (see Fig. \ref{cumu}). It can be seen that, like in
observations, the dark 
remnant mass is significantly more concentrated than all mass groups in 
simulation 128kN7, while in other simulations it broadly overlaps the 
distribution of the most massive luminous stars. This difference likely arises
from the larger efficiency of mass segregation in simulation 128kN7 with
respect to the other two (see Fig. 3 of Contenta et al. 2015). In fact, in this 
simulation, the difference between the mean mass of WDs and MS stars produces a 
significant central segregation of the former.

\begin{figure}
 \includegraphics[width=8.6cm]{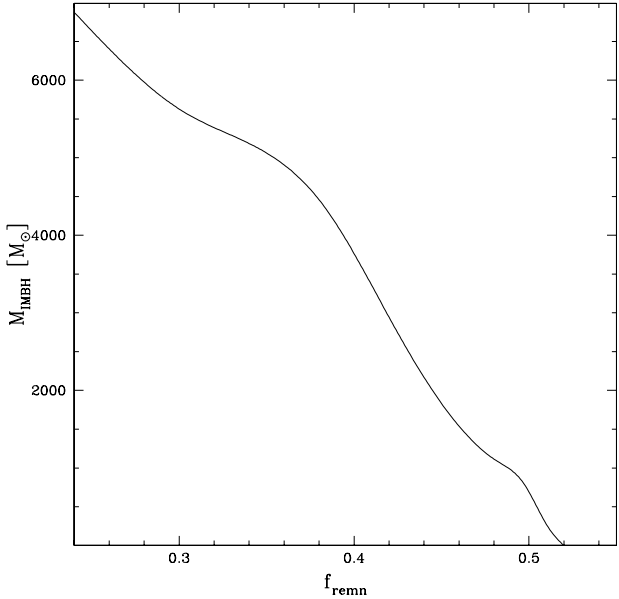}
 \caption{Best-fit IMBH mass in NGC 6218 as a function of the adopted global mass fraction of
 dark remnants.}
\label{imbh}
\end{figure}

Alternatively, it is also possible that part of the mass excess estimated in the
two analysed GCs is
concentrated in a single massive remnant located in the cluster center i.e. an
intermediate-mass BH (IMBH). Indeed, the method described in Sect. \ref{met_dyn_sec} is
not able to resolve density variations within the innermost pc where only a few
radial velocities are available. The mass of such an IMBH also depends on
the fraction and distribution of other remnants. To test this hypothesis we
adopted the technique described in Sollima et al. (2012) to simultaneously fit the surface
brightness profile and the core luminosity function of NGC 6218
with multimass King-Michie models (Gunn \& Griffin 1979) with different 
fractions of dark remnants. A point mass potential has been then added to the 
model potential to simulate the effect of an IMBH and the LOS velocity dispersion
has been derived by integrating eq. \ref{jeans_eq} and using the projection
formula in eq. \ref{mge2_eq}. Of course, the addition of the point-mass could
in principle lead to solutions that are not fully self-consistent. On the other hand,
given that $M_{IMBH}<<M_{dyn}$, this effect is expected to be small and should
not affect the following considerations.
The best-fit IMBH mass has been chosen to maximize the log-likelihood 
(eq.\ref{sig_eq2}). In Fig. \ref{imbh} the best-fit IMBH mass is plotted as a
function of the adopted remnants global mass fraction. As expected, the larger is the
fraction of remnants ($f_{remn}$) the smaller is the mass of the hypothetical IMBH. Note that
for any dark remnants mass fraction $>$24\% (a strong lower limit according
to the considerations made above) the IMBH mass is always $<7\times10^{3}
M_{\odot}$ while for $f_{remn}>52\%$ no IMBH are allowed.

\section{Summary}

The main result obtained in this paper is that the largest fraction ($>60\%$) of
the mass content of both the analysed GCs is dominated by centrally concentrated
non-luminous mass.

The results presented here confirm what already found in Sollima et al. (2012).
However, a significant improvement with respect to that work has been made here
since: {\it i)} the adopted dataset of radial velocities is $\sim$5 times larger
within the half-light radii of both clusters, {\it ii)} the
analysis presented here is model-independent as it requires no
assumptions on the degree of mass segregation among mass groups, and {\it iii)}
in the present work we are able to determine the radial distribution of the dark
mass.

The obtained result seems not to be due to the heating produced by tidal effects 
or binary stars being therefore linked to a real excess of dark mass in the
central region of these clusters.

The most likely hypothesis is that such a large mass excess is produced by dark
remnants sunk in the cluster core during the cluster lifetime as a result of
mass segregation. In fact, the comparison with N-body simulations indicates that GCs
losing a significant ($>70\%$) fraction of their initial mass contain a
centrally concentrated core of dark remnants constituted mainly of WDs accounting
for $\sim$50\% of the total cluster mass within the same radial extent explored
in the present analysis. This value is slightly lower, but within the
uncertainties, than that estimated in our analysis.
It is worth noting that the N-body simulations considered in this work were not
specifically run to reproduce the present-day structure of the two analysed 
clusters and are subject to a different tidal stress. 
So, only qualitative conclusions can be
drawn from such a comparison. 
However, similar results are obtained by other authors: 
Giersz \& Heggie (2011) used Monte Carlo simulations with
$2\times10^{6}$ particles to model the present-day structure of 47 Tuc and found
that the fraction of WDs is $\sim51\%$ in the core and even larger in a scaled
simulation run with a flatter initial mass function. Sippel et al. (2012) and
L{\"u}tzgendorf et al. (2013)
analysed a set of N-body simulations including
natal kicks for NSs and BHs and found that after 12 Gyr $\sim40\%$ of the
cluster mass is contained in WDs (and a larger fraction is expected in the core).
According to Vesperini \& Heggie (1997) and Baumgardt \& Makino (2003) the 
fraction of WDs and NSs steadly increases as a cluster evolves and loses
mass, becoming dominant when the cluster lost $>60\%$ of its stars. 
In this regard, a large mass-loss rate in the two analysed clusters is also 
suggested by their relatively flat mass functions (de Marchi, Pulone \& Paresce
2006; Paust et al. 2010; Sollima et al. 2012).
On the basis of the above consideration, it is still possible that the observed 
overabundance of non-luminous mass in the two GCs analysed here is entirely due 
to a compact population of centrally segregated WDs.
A valuable test to this hypothesis would be to measure the number of WDs in the
central region of these clusters through photometric analyses. Unfortunately, to
perform a complete census of WDs in these clusters it would be necessary to reach
extremely faint magnitudes ($V\sim30-31$) with a good level of completeness, an
unfeasible task even with HST.

The possibile presence of an IMBH in the center of these two GCs, as 
suggested for many other Galactic GCs by
L{\"u}tzgendorf et al. (2013; but see van der Marel \& Anderson 2010 and 
Lanzoni et al. 2013), cannot be excluded. In particular, according to
the actual fraction of retained remnants, the mass of such an hypothetical BH
could lie in the range $M_{IMBH}<7\times10^{3} M_{\odot}$. On the other hand, we believe that given
the above apparent degeneracy and the large uncertainty in the expected dark
remnants fractions, any claim of IMBH detection is inappropriate. Similarly, the
presence of an IMBH can significantly affect the determination of the actual 
fraction of remnants.

Another possibility is that the mass excess observed here is due to a modest
amount of non baryonic DM. 
Previous investigations in this regard made on the outer halo GC NGC 2419
seem to rule out the presence of a significant amount of DM within the stellar extent of this
cluster (Baumgardt et al. 2009; Conroy, Loeb \& Spergel 2011) although the
involved uncertainties in the anisotropy profile leave some room for a
small DM content (Ibata et al. 2013).
Note that in this scenario the DM halo should
extend far beyond the extent of the stellar component. N-body simulations 
assuming GCs surrounded by cored DM halos predict that, because of the 
interaction with the cluster stars (Baumgardt \& Mieske 2008) and stripping
by the tidal field of the host galaxy, the central parts of
GCs might be left relatively poor in DM because
at the present time the DM either populates the outer region of the
cluster or has been stripped (Mashchenko \& Sills
2005; Pe{\~n}arrubia, Navarro \& McConnachie 2008).
In this scenario, the high concentration of dark mass evidenced in
our analysis would favor a cuspy shape of the surviving DM halo dominating the
cluster mass budget in the central region of these clusters.
This is however in contrast with theoretical considerations (where 
interactions with baryons are expected to remove DM cusps; Navarro, Eke \& Frenk 
1996; Mo \& Mao 2004; Mashchenko, Couchman \& Wadsley 2006; Del
Popolo 2009; Di Cintio et al. 2014; Pontzen \& Governato 2014; Nipoti \& Binney 2015). 

Because of the limited sample of GCs analysed here it is not possible to
check the presence of correlation between dark mass content and other dynamical
and general parameters. Future studies addressed to the extension of this
analysis to a larger sample of GCs will help to clarify the nature of the dark
mass and the effect of the various dynamical processes occurring during the
cluster evolution on its fraction. 

\section*{Acknowledgments}

AS acknowledges the PRIN INAF 2011 ``Multiple populations in globular
clusters: their role in the Galaxy assembly" (PI E. Carretta) and the PRIN INAF
2014 ``Probing the internal dynamics of globular clusters. The first,
comprehensive radial mapping of individual star kinematics with the
generation of multi-object spectrographs" (PI L. Origlia). 
AS, FRF, EL, BL, AM, ED and CP acknowledge support from the European 
Research Council (ERC-2010-AdG-267675, COSMIC-LAB).
FC acknowledge support from the European Research Council (ERCStG-335936, CLUSTERS).
We warmly thank Michele Bellazzini, Yazan Momany, Anna Lisa Varri and Douglas 
Heggie for useful discussions and suggestions. We also thank the anonymous
referee whose helpful comments and suggestions improved our paper.

\appendix
\onecolumn

%
%

\section{Distribution function from Multi-Gaussian expansion of density and
velocity dispersion profiles}

According to Eddington (1916), the distribution function of a spherical and
isotropic system
can be derived from the formula
\begin{eqnarray}
f(E)&=&-\frac{1}{\sqrt{8}\pi}\left[\int_{E}^{\phi_{t}}\frac{d^{2}\rho}{d\phi^{2}}\frac{d\phi}{\sqrt{\phi-E}}-\frac{1}{\sqrt{\phi_{t}-E}}\left(\frac{d\rho}{d\phi}\right)_{\phi=\phi_{t}}\right]\nonumber\\
&=&-\frac{1}{\sqrt{8}\pi}\left[G\int_{r(\phi=E)}^{r_{t}}\frac{d^{2}\rho}{d\phi^{2}}\frac{M(r)}{r^{2}\sqrt{\phi(r)-E}}d
r-\frac{1}{\sqrt{\phi_{t}-E}}\left(\frac{d\rho}{d\phi}\right)_{\phi=\phi_{t}}\right]
\label{edd_eq}
\end{eqnarray}
where $\phi_{t}=-GM/r_{t}$ is the potential at the cluster tidal radius.
The above equation can be generalized to account for Osipkov-Merritt anisotropic models by 
replacing $E$ with
$$Q\equiv \phi_{t}-E-L^{2}/2r_{a}^{2}$$
and $\rho$ with
$$\rho_{Q}\equiv(1+\tilde{r}^{2})\rho$$
Any physically meaningful model must satisfy the inequality $f(Q)>0$ across the
entire range of $Q$. 

All the terms in eq. \ref{edd_eq} can be written as a function of the 3D mass, 
global density profile (eq. \ref{rho_eq}) and its radial derivatives.
\begin{eqnarray*}
\frac{d\rho_{Q}}{d\phi}&=&\frac{r}{G
M(r)}\left[r (1+\tilde{r}^{2})\frac{d\rho}{d r}+2\rho\tilde{r}^{2}\right]\nonumber\\
\frac{d^{2}\rho_{Q}}{d\phi^{2}}&=&\left(\frac{r}{G
M(r)}\right)^{2}\left[r^{2}(1+\tilde{r}^{2})\frac{d^{2}\rho}{d r^{2}}+2 r
(1+3\tilde{r}^{2})\frac{d\rho}{d r}-\frac{4\pi\rho
r^{4}}{M(r)}(1+\tilde{r}^{2})\frac{d\rho}{d
r}+2\rho\tilde{r}^{2}\left(3-\frac{4\pi\rho
r^{3}}{M(r)}\right)\right]\nonumber\\
\phi(r)&=&-G\left[\frac{M(r)}{r}+4\pi\int_{r}^{r_{t}}\rho~r~dr\right]
\end{eqnarray*} 

So, the distribution function $f(Q)$ can be derived by numerically integrating 
eq. \ref{edd_eq} for
each value of $Q$ in the range $\phi(0)<Q<\phi_{t}$.

\label{lastpage}

\end{document}